\documentclass[showpacs,preprintnumbers,preprint,nofootinbib,superscriptaddress,showkeys]{revtex4}

\usepackage{amssymb}
\usepackage{amsmath}
\usepackage{graphicx}
\usepackage{dcolumn}
\usepackage{bm}
\usepackage{epsf}

\def\be{\begin{eqnarray}}
\def\ee{\end{eqnarray}}
\def\bea{\begin{eqnarray}}
\def\eea{\end{eqnarray}}

\begin{document}


\title{Breakdown of the universality in few-boson systems}
\author{Arnoldas Deltuva}
\email{deltuva@cii.fc.ul.pt}
\affiliation{Centro de F\'{\i}sica Nuclear da Universidade de Lisboa,
P-1649-003 Lisboa, Portugal }
\author{Rimantas Lazauskas}
\email{rimantas.lazauskas@ires.in2p3.fr}
\affiliation{IPHC, IN2P3-CNRS/Universit\'e Louis Pasteur BP 28,
F-67037 Strasbourg Cedex 2, France}

\date{\today }
\pacs{36.40.-c,34.50.-s,21.45.-v}

\begin{abstract}
We develop a series of resonant short-range two-boson potentials reproducing the same
two-body low-energy observables and apply them in three- and four-body calculations.
We demonstrate that the
universal behavior predicted by effective field theory may be strongly violated
and analyze the conditions for this phenomenon.

\end{abstract}

\maketitle

\renewcommand{\thefootnote}{\#\arabic{footnote}} \setcounter{footnote}{0}

\section{Introduction}


Effective field theory (EFT) found a great success in particle, nuclear, and
condensed matter physics. EFT and the alternative quantum-mechanical approach
with EFT-generated interactions turn out to be a very practical tools to deal with
the problems involving largely separated energy (or length) scales. One such
example, revealed by the nuclear and cold atomic gas structures,
is a system of interacting particles at low energies whose behavior is governed by the large
two-particle scattering length $a_0$, i.e., $|a_{0}| \gg r_0$, $r_0$ being the range
of the interaction, while by
low energy we understand that characteristic center-of-mass
(c.m.) momenta $k$ are significantly smaller than $1/r_{0}$).
In this case, in order to describe the occurring physical phenomena,
 one naturally introduces EFT expansion
in small dimensionless parameters $r_{0}/a_{0}$ and $kr_{0},$ which should be
well behaved apart from the well-known case of singular zero-range interaction
($r_{0}\rightarrow 0$), leading to Thomas collapse in the $N>2$
particle systems with attractive pairwise interactions.

The  $r_{0}>0$ case has been first seriously investigated by Efimov \cite{Efimov_70,Efimov_79},
who has derived several interaction-independent scaling relations for
three-particle observables of the systems with the particles  constrained
in the region where their interdistance $R$ satisfies the relation
$r_{0} \ll R \ll |a_{0}| $. Notably, as the
ratio $a_{0}/r_{0}$ increases,  the
accumulation of the weakly bound three-particle states (Efimov states)
 has been demonstrated.
Efimov states have found much attention
recently. Properties of the simplest many-particle structures, namely $0^{+}$
states in the systems of identical bosons, have been classified. In
particular, numerous universal (interaction independent) relations have been
established between three-boson observables. It has been accepted
that in the first approximation (leading order of EFT) all the low-energy
properties of three-boson system are set by two parameters: one two-body
parameter --- two-boson (dimer)  binding energy or scattering length, and one
three-body parameter ---  three-boson (trimer) binding energy or particle-dimer scattering
length~\cite{MC_HAMMER,Frederico_02}. More recently, it has even been demonstrated that properties of
the four-boson system (tetramer) are governed by the same two parameters and no
additional four-body scale is required to establish universal relations
between three- and four-boson observables~\cite{Platter,stecher}.  Universality
studies have been extended also to fermionic three- and four-body systems,
by means of the so-called pionless EFT, recovering phenomenologically
observed Phillips (triton binding energy vs. neutron-deuteron doublet scattering length)
and Tjon  (triton vs. alpha particle binding energies) correlations \cite{Platter_N}.
 Most of these relations have been
derived relying on the purely attractive (and often contact) two-particle interactions, while $N>2$
 systems are balanced to prevent Thomas collapse by
introducing repulsive three-particle force.
Such approach evidently helps to preserve
Efimov relation $r_{0}\ll R\ll | a_{0}| $ in all
generated structures. Nevertheless, in real physical systems, if the
two-particle cluster is resonant,  one can guarantee only $R\ll \left\vert
a_{0}\right\vert $ condition but one cannot always
assure that three- or four-particle system prevails the condition $r_{0}\ll R$,
i.e., most of the time particles stay outside the interaction region in the
few-particle system.

In this work we follow the quantum-mechanical approach with slightly
different interactions, namely, we
develop series of short-range separable rank 2 potentials  preserving the same
two-boson binding energy and low-energy scattering observables.
These potentials have
been derived somehow mimicking interaction between $^{4}$He atoms, i.e.,
choosing the boson mass  $\hbar^{2}/m=$ 12.12 K$\cdot${\AA}$^{2}$
and fixing the two-boson scattering length to $a_{0}=$104.0 {\AA}, but we admit that
a realistic description of  $^{4}$He multimers is not the aim of the present work.
Instead, with these potentials we calculate and compare series of three- and four-particle
properties, including binding energies, mean square radii, and scattering lengths.
We demonstrate that, despite resonant two-boson interaction,
the breakdown of the  universality and commonly established EFT results is possible,
and investigate conditions for the occurrence of this phenomenon.

In Section \ref{sec:dyn} we describe the dynamic input and recall the methods used to solve
few-particle equations. In Section \ref{sec:res} we present the results for three- and
four-boson bound states and scattering lengths.
Section  \ref{sec:con} contains our conclusions.

\section{Dynamics \label{sec:dyn}}

As a dynamic input to our calculations we  use a set of two-boson potentials
that reproduce the same two-particle binding energy and low-energy scattering observables.
Such a set could be generated from the initially chosen potential $v^{(0)}$
by unitary transformations
like those of the renormalization group \cite{bogner}. The resulting potentials
are nonlocal even if the initial one was local. We follow a simpler procedure and
use separable (and thereby nonlocal) potentials
\begin{equation}
v = \sum_{ij}^{n_r} |g_i\rangle \lambda_{ij} \langle g_j|
\end{equation}
that, for simplicity, are restricted to act in the $S$-wave only. In close analogy with the most
EFT-generated potentials~\cite{Platter}, the form factors $|g_i\rangle$  are chosen as Gaussians
that in the momentum-space representation are defined as
\begin{equation}
\langle k |g_i\rangle = e^{-(k/\Lambda_i)^2}
\end{equation}
whereas in the configuration-space they are
\begin{equation}
\langle r |g_i\rangle = (\Lambda_i/\sqrt{2})^3 \, e^{-(\Lambda_i r/2)^2}.
\end{equation}
The initial potential $v^{(0)}$ was chosen to be of rank one,
i.e., $\lambda_{ij}^{(0)}=\delta_{i1}\delta_{j1} \lambda_{11}^{(0)}$,
with the momentum cutoff $\Lambda_1^{(0)} = 0.2$ \AA$^{-1}$
and $\lambda_{11}^{(0)} = [\pi m (1/a_0 - \Lambda_1^{(0)}/\sqrt{2\pi})/2]^{-1}$
determined by the two-body scattering length  $a_{0}=$104.0 {\AA}.
The corresponding dimer binding energy and the
effective range are $B_2 = 1.318$ mK and $r_0 = 15.0$ \AA, respectively;
the latter is about two times larger than the value provided by the
"realistic" interaction models for $^{4}$He atoms.
All the other potentials were obtained by choosing $\Lambda_j = j\Lambda$
with $\Lambda$ values in the range between 0.133 and 0.305 \AA$^{-1}$,
i.e., $13.8 < a_0 \Lambda < 31.7$,
and determining  the respective strength parameters $\lambda_{ij}$
from the fit of the calculated observables, i.e.,
 $B_2$,  $a_{0}$, and two-particle scattering phase-shifts up to about $50 \, B_2$ c.m. energy,
 to the predictions of the initial potential $v^{(0)}$;
we found that rank $n_r=2$ potential with three free parameters $\lambda_{11}$,
$\lambda_{22}$, and $\lambda_{12} = \lambda_{21}$ is sufficient for
a high quality fit with a typical  four digit accuracy.
Furthermore, a broad interval of the scattering energies  included in the fit,
roughly corresponding to the natural energy scale   $\hbar^{2}/mr_0^2 $ \cite{Platter},
guarantees that in the effective range expansion not only $k^2$ but also
higher order terms are  well retained.

The separable form of the potential is convenient for the fit but the advantage of
the separability is not used when solving few-body equations
that can include any form of the potential.
The exact three- and four-body bound state and scattering
equations are solved numerically by two completely
different methods. The symmetrized form of equations is employed which is appropriate
for the system of identical bosons.
The configuration-space Faddeev and Faddeev-Yakubovsky (F/FY)
integro-differential equations \cite{Fad,Yakub}
for the wave function components are solved using the numerical technique of
Ref.~\cite{FY_Gre}.
In the momentum-space framework the integral form of F/FY equations
is used for the description of the  bound states
whereas  in the case of scattering the
Alt, Grassberger, and Sandhas  (AGS) equations \cite{alt:67a,grassberger:67}
for the transition operators  (see  Ref.~\cite{deltuva:07c} for the symmetrized version)
leading more directly to the observables
are solved following the technical developments of Refs.~\cite{deltuva:03a,deltuva:07a}.
Both calculations are performed in the partial wave basis. Although
the two-body interaction is limited to the $S$ wave, higher angular momentum
states up to $l_y, \, l_z \leq 2$ are needed for the particle-pair and particle-triplet
relative motion. With those states included we obtain
 well converged results, while both methods
are in excellent agreement, the differences being well below 0.1\%.

\section{Results \label{sec:res}}

\begin{figure}[tbp]
\epsffile{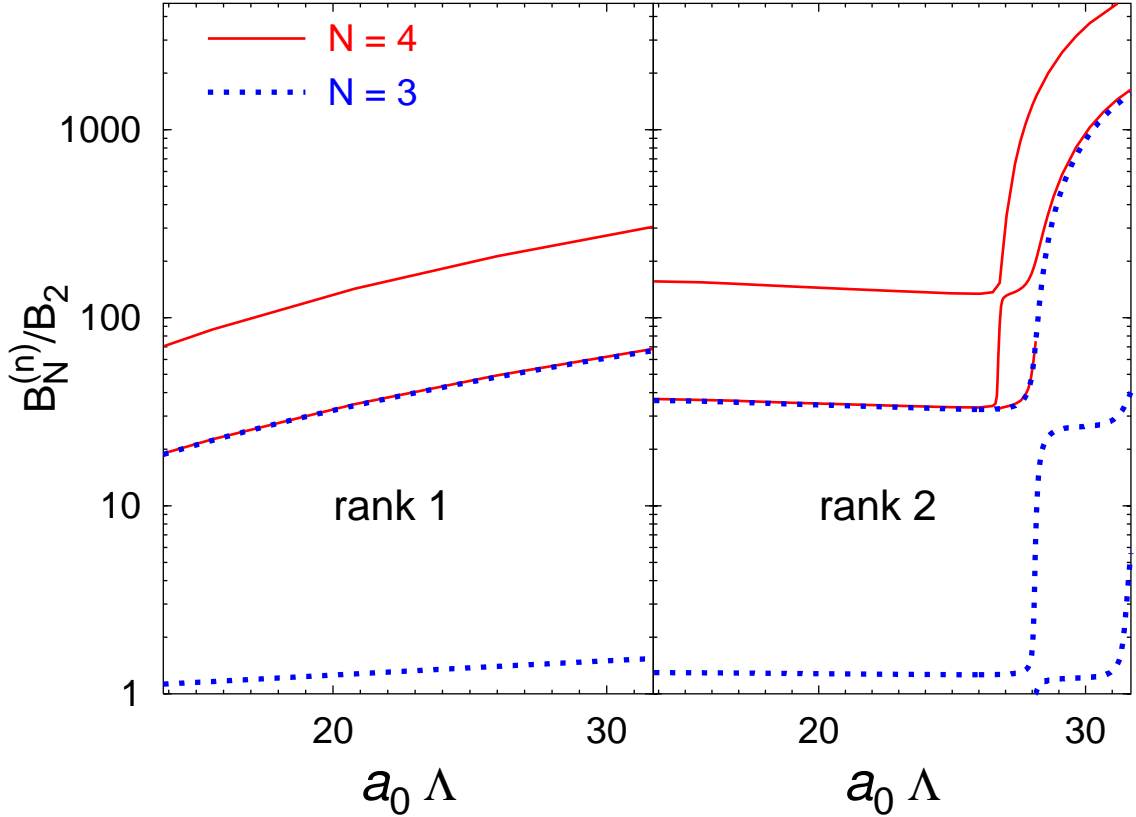}
\caption{\label{fig:b-l}
(Color online)
Evolution of the three- and four-boson binding energies with the cutoff parameter $\Lambda$.}
\end{figure}

We would like to note first that   $^{4}$He multimers
in the quantum-mechanical approach with EFT-generated potentials
have been also studied by Platter et al.~\cite{Platter}.
However, rank $n_r=1$ potentials have been used in the latter work
with $\lambda_{11}$ fitted to the dimer binding energy $B_2$;
although in the leading order in $(a_0 \Lambda)^{-1}$ the relation
$a_0=[\hbar^2/(mB_2)]^{1/2}$ holds,  strictly speaking obtained $a_0$  as well as the
scattering phase shift values are cutoff-dependent in this approach.
To renormalize the three-body system,
i.e., to get (nearly) cutoff-independent results for physical three-body observables,
an additional separable three-boson force (repulsive or attractive, depending
on  the $\Lambda$ value) had to be included in the calculations of Ref.~\cite{Platter};
otherwise, the three- and four-boson binding energies show quite strong dependence
on $\Lambda$ which is interesting to compare with our rank 2 results.
We perform   $n_r=1$ calculations with  $\lambda_{11}$ fitted to the  same
two-boson scattering length $a_0 = 104.0$ \AA {} as in the $n_r=2$ case.
The two approaches provide very different $\Lambda$-evolution
of the few-boson binding energies as  Fig.~\ref{fig:b-l} demonstrates.
The ratios of trimer and tetramer binding energies $B_N^{(n)}$ to the dimer binding
energy $B_2$
for  the rank 1 potentials increase monotonically with  $\Lambda$,  whereas for  rank 2 they
show much weaker  dependence in the  range $a_0\Lambda < 26.0$
where all $B_N^{(n)}/B_2$ slowly decrease with  $\Lambda$.
In that region all potentials support ground and one excited state for trimer and tetramer,
whereas the  tetramers excited state binding energy $B_4^{(1)}$ is only slightly larger than
$B_3^{(0)}$ and thus follows closely the trimers ground state curve;
the same applies to the $n_r=1$ case in the whole considered region.
From  rank 2 calculations in the  range $a_0\Lambda < 26.0$  it may seem that when the
two-boson observables are fixed by a proper fit there may be no need for a three-body
parameter to get (nearly) cutoff-independent results for the three-body and four-body system.
However, for larger $\Lambda$ values
all $B_N^{(n)}/B_2$  in the $n_r=2$ case show a nontrivial behavior despite the fact
that all potential parameters $\lambda_{ij}$ vary smoothly and slowly with $\Lambda$.
First, at $a_0\Lambda = 26.5$  the tetramers excited state binding energy $B_4^{(1)}$
detaches from $B_3^{(0)}$ and exhibits a very rapid increase but stabilizes for a moment
when it almost attains
the tetramer ground state binding energy $B_4^{(0)}$ that, with a tiny delay, also
starts to increase rapidly.
Furthermore, at $a_0\Lambda = 26.75$ the second excited tetramer state appears with the
binding energy $B_4^{(2)}$ only slightly larger than $B_3^{(0)}$.
In this region of drastic tetramer variations the trimer properties remain practically
unchanged, suggesting that not a three- but four-body parameter would be needed for the renormalization
of the considered few-particle system~\cite{Frede_4b}.
The need of the four-body parameter as well as the existence of three tetramer
bound states disagrees with the universality predictions \cite{Platter,stecher};
the reason for this nonstandard behavior will be explained later.
The trimer also shows  qualitatively similar behavior just  at larger $\Lambda$ values.
 Around $a_0\Lambda = 27.5$ the binding energies of the trimer ground and first excited state,
 $B_3^{(0)}$ and $B_3^{(1)}$, respectively, start to grow very rapidly; the former increases much like $B_4^{(0)}$
while the latter stabilizes for a while.
As  $B_3^{(0)}$  increases the second excited tetramer state disappears at  $a_0\Lambda = 28.14$
and the  binding energy $B_4^{(1)}$ of the first excited tetramer state comes close to $B_3^{(0)}$ again.
Furthermore, at $a_0\Lambda = 28.12$ the second excited trimer state appears.
Unlike the stronger bound trimers, this one
has properties of the true Efimov state, i.e., it
slides under the dimer threshold if the two-boson interaction is made stronger.
The second excited trimer state loses that property above $a_0\Lambda = 30.7$
where its binding energy  $B_3^{(2)}$, much like  $B_3^{(1)}$  at $a_0\Lambda = 27.5$,
enters a phase of a rapid increase.
However, the Efimov property is recovered in the third excited trimer state
appearing at $a_0\Lambda = 31.7$.  It is interesting to note
that the second excited tetramer state shows a similar behavior, i.e., it also
disappears if the two-boson interaction is made stronger.
Furthermore, at  $\Lambda$ values corresponding to
the appearance of the second and third excited trimer states the binding energy of a deeper
lying trimer state is almost the same in both cases,  about $5.6 B_2$.
This value is slightly smaller than one found 
 in Ref.~\cite{Frederico_02} for zero-range potential, which should be consequence of the
 finite range.
As can be seen from  Fig. 31 of Ref.~\cite{MC_HAMMER}, zero-range calculations
with fixed  $B_2$ and $B_3^{(n)}$ slightly underestimate $B_3^{(n+1)}$ values
compared to the results obtained with realistic potentials.
Therefore the critical value of $B_3^{(n)}/B_2$ is lower for the finite-range
potentials.

\begin{figure}[tbp]
\epsffile{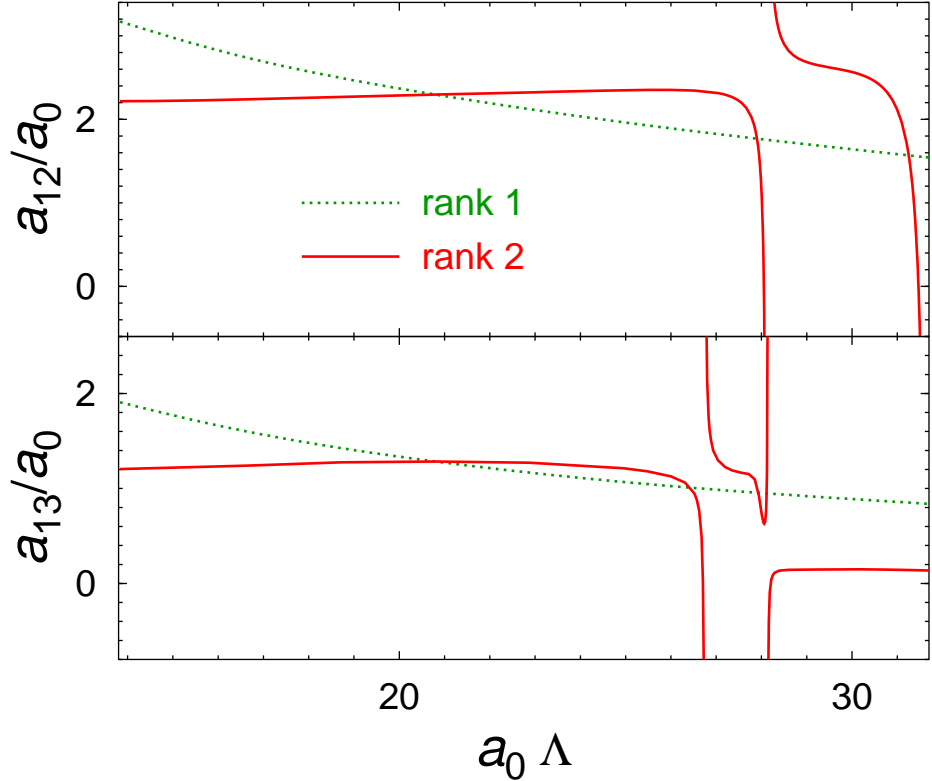}
\caption{\label{fig:a}
(Color online)
Evolution of the  atom-dimer and atom-trimer scattering length
 with the cutoff parameter $\Lambda$.}
\end{figure}

The appearance/disappearance of trimer and tetramer excited states can be pointed-out
in the $\Lambda$-dependence of the atom-dimer and atom-trimer scattering lengths
$a_{12}$ and $a_{13}$ shown in Fig.~\ref{fig:a}:
when a new  trimer (tetramer) state appears, $a_{12}$ ($a_{13}$) has a discontinuity
going to $-\infty$ and returning from $+\infty$. Thus,  Figure~\ref{fig:a} demonstrates
that new trimer states appear around $a_0\Lambda = 28.12$ and 31.7, whereas an additional
tetramer state exists only for $26.75 < a_0\Lambda < 28.14$.
Much like the binding energies, the scattering lengths calculated with the rank 2 potentials
show only weak $\Lambda$-dependence at $a_0\Lambda < 26.0$, whereas more significant but monotonic
$\Lambda$-dependence is seen in the  rank 1 case.

\begin{figure}[tbp]
\epsffile{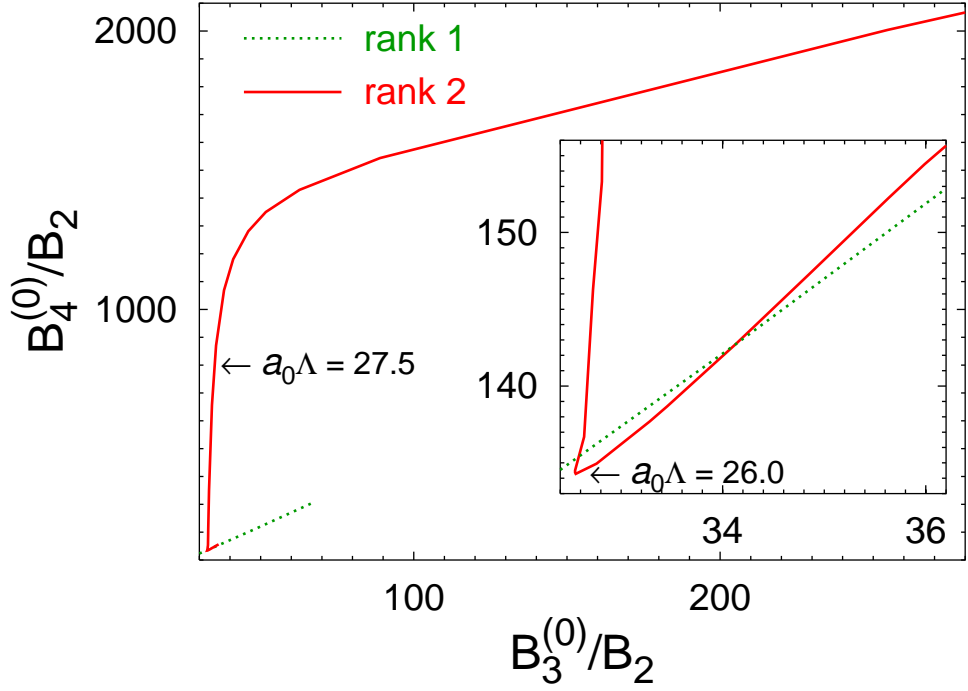}
\caption{\label{fig:b4}
(Color online)
Correlation between trimer and tetramer ground state binding energies.}
\end{figure}

The EFT predicts universal nearly linear correlations between the
three and four-particle binding energies~\cite{Platter}, revealed
by the well-known Tjon line in the few-nucleon
physics.  As example we show in Fig.~\ref{fig:b4} the relation
between trimer and tetramer ground state binding energies.
Results obtained with $n_r=1$ potentials are roughly consistent with the findings of EFT in the
whole considered $\Lambda$ region where $B_4^{(0)}/B_2$ as a function of $B_3^{(0)}/B_2$
has slope around 5. However, our $n_r=2$ results obviously violate
linear correlations around  $a_0\Lambda = 27.5$ where the three-boson binding energies  $B_3^{(n)}$
are almost $\Lambda$-independent, whereas the four-boson binding energies  $B_4^{(n)}$
exhibit a very rapid increase.
There is even a very narrow region around $a_0\Lambda = 26.0$
where trimer binding energies decrease, while at
the same time tetramer binding energies grow with $\Lambda$.
Nevertheless, nearly linear correlations are preserved
at  $a_0\Lambda < 26.0$, also for the excited states.
The linear correlation but with a different slope, around 2.4,
takes place also at  $a_0\Lambda > 28.5$ where both $B_3^{(0)}$ and  $B_4^{(0)}$ increase rapidly;
however, due to small radii and high average kinetic energies, as will be shown later,
this is not the regime where one should expect validity of the universal relations.

\begin{figure}[tbp]
\epsffile{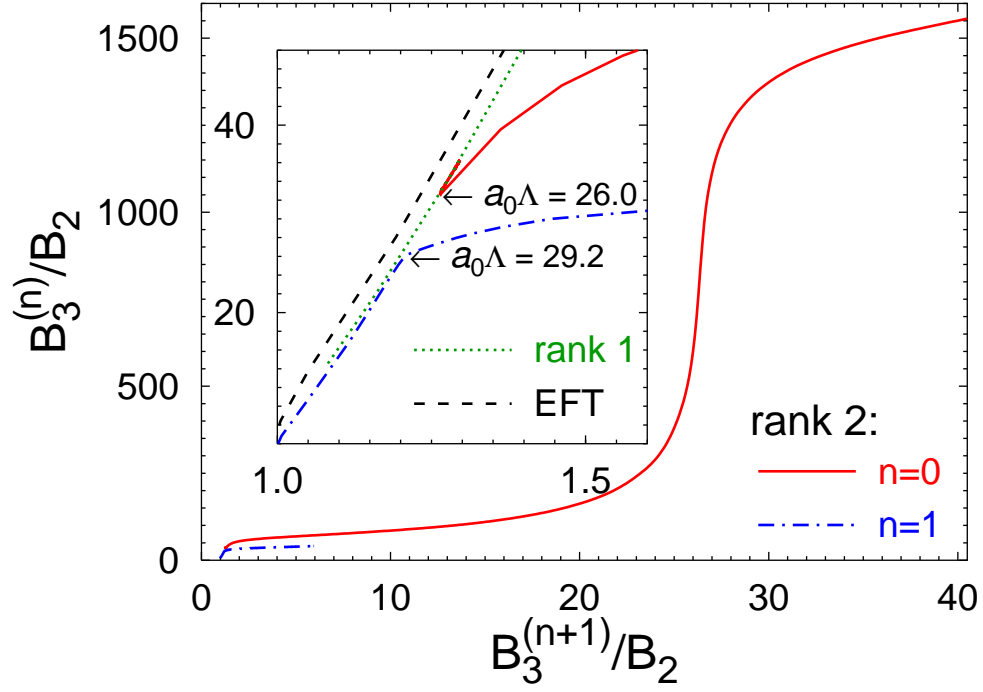}
\caption{\label{fig:b3}
(Color online)
Correlations of the trimer binding energies.}
\end{figure}

Correlations between trimers ground and excited state binding energies
are shown in Fig.~\ref{fig:b3}.
In contrast to the nontrivial behavior of the results obtained with rank 2 potentials,
contact interaction EFT, which ensures
validity of the Efimov condition, proclaims universality of the $B_3^{(n)}/B_2(B_3^{(n+1)}/B_2)$ curve
\cite{MC_HAMMER}. The results obtained with realistic ${}^4$He potentials lie slightly
below the universal curve \cite{MC_HAMMER} much like our rank 1 results do.
On the other hand, the relation between $B_3^{(0)}$ and $B_3^{(1)}$ obtained with
$n_r=2$ potentials agrees well with the rank 1 curve only at $a_0\Lambda < 26.0$
where binding energies decrease by increasing $\Lambda$.
At $a_0\Lambda = 26.0$ the  $B_3^{(0)}/B_2(B_3^{(1)}/B_2)$ function exhibits a critical point
behavior and recedes from the  rank 1 curve.
The second excited trimer state appears at $a_0\Lambda = 28.12$;
the corresponding correlation function $B_3^{(1)}/B_2(B_3^{(2)}/B_2)$ increases monotonically
and is consistent with the rank 1 $B_3^{(0)}/B_2(B_3^{(1)}/B_2)$ curve
till $a_0\Lambda = 29.2$ but then deviates from it as well.

\begin{figure}[tbp]
\epsffile{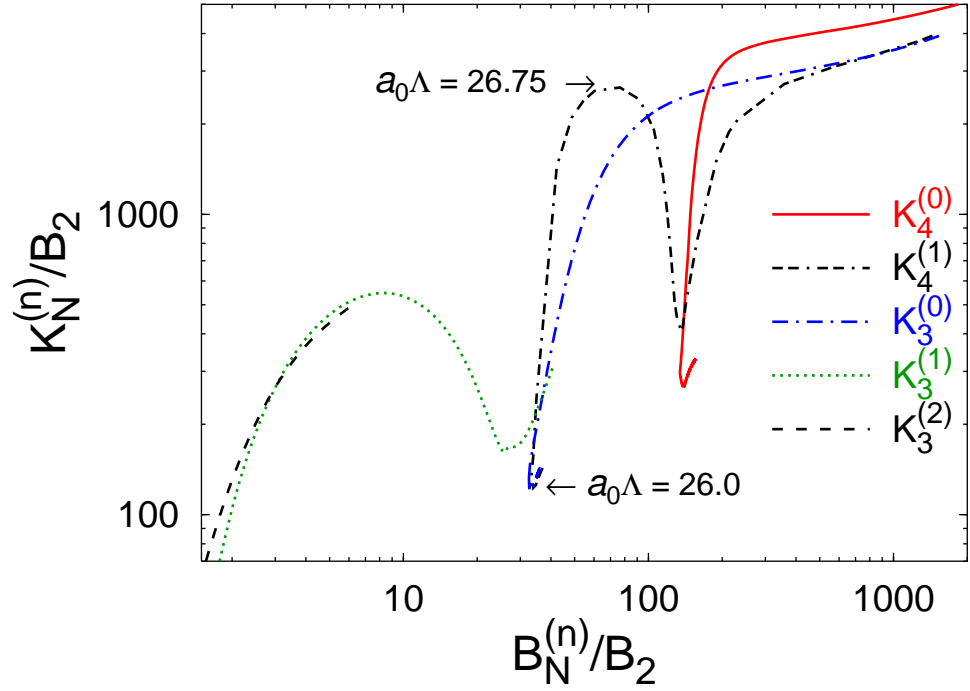}
\caption{\label{fig:K}
(Color online)
Correlations between the expectation values of kinetic energy and
the corresponding binding energies.}
\end{figure}

Given such a nontrivial behavior of few-boson binding energies it is worth to study
other bound state properties such as the expectation value of the kinetic energy operator,
$K_N^{(n)}$, and the mean square radius, $R_N^{(n)}$.
As shown in Fig.~\ref{fig:K}, there is at least a qualitative correlation between
kinetic and respective binding energies  of the trimer and tetramer ground state,
i.e., except for the vicinity of the critical points,  $K_N^{(0)}$ increase with $B_N^{(0)}$,
although the slope varies quite strongly.
Evolution of the  first excited state kinetic energies is more peculiar.
In the region $a_0\Lambda > 26.0$, where binding energies start to grow rapidly with $\Lambda$,
the phase of the fast increase of kinetic energies $K_N^{(1)}$  is followed by the phase of
fast decrease after which the  $K_N^{(1)}$ increases in a way similar to $K_N^{(0)}$.

\begin{figure}[tbp]
\epsffile{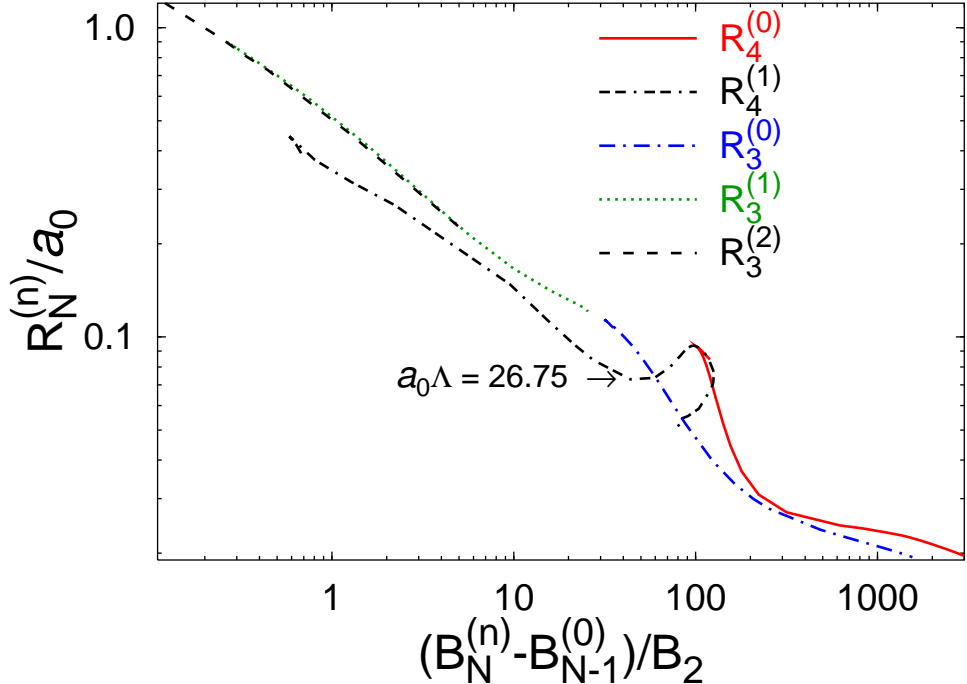}  
\caption{\label{fig:R}
(Color online)
Correlations between the mean square radii and
the corresponding separation energies.}
\end{figure}

The correlations between mean square radii  $R_N^{(n)}$ and the corresponding
separation energies $[B_N^{(n)}-B_{N-1}^{(0)}]$ are shown in  Fig.~\ref{fig:R}.
We note that for the ground states  $R_N^{(0)} < r_0 = 0.144 \, a_0$, thus, the observed
deviations from the universal behavior are not surprising.
As it is natural to expect, the radii decrease with increasing separation energy,
often the dependence is not far from  linear. The exception is the first excited state
of the tetramer at $a_0\Lambda > 26.75$.
{The origin of this peculiar behavior is unveiled in the next paragraph, when discussing
the shape of non-local potential.}
Worth noting is that in a small window around $a_0\Lambda = 26.75$, corresponding to the
first rapid increase of $B_4^{(1)}$, the
first $K_4^{(1)}$  peak and the first $R_4^{(1)}$ minimum, the tetramers excited state kinetic
energy is larger and its mean square radius is smaller that the corresponding quantities of
the ground state, i.e.,  $K_4^{(1)} > K_4^{(0)}$ and  $R_4^{(1)} <  R_4^{(0)}$.

\begin{figure}[tbp]
\begin{center}
\mbox{\epsfxsize=11.9cm\epsffile{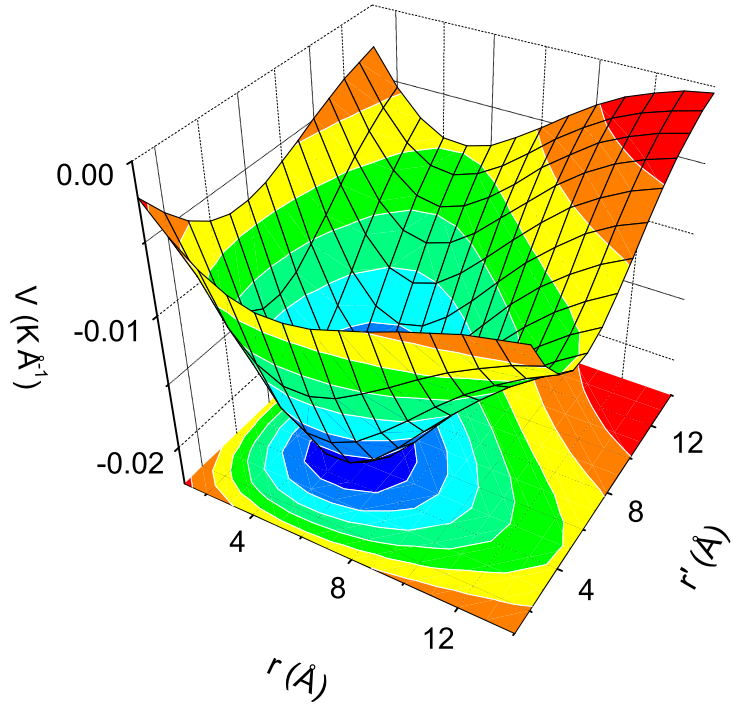}} %
\mbox{\epsfxsize=11.9cm\epsffile{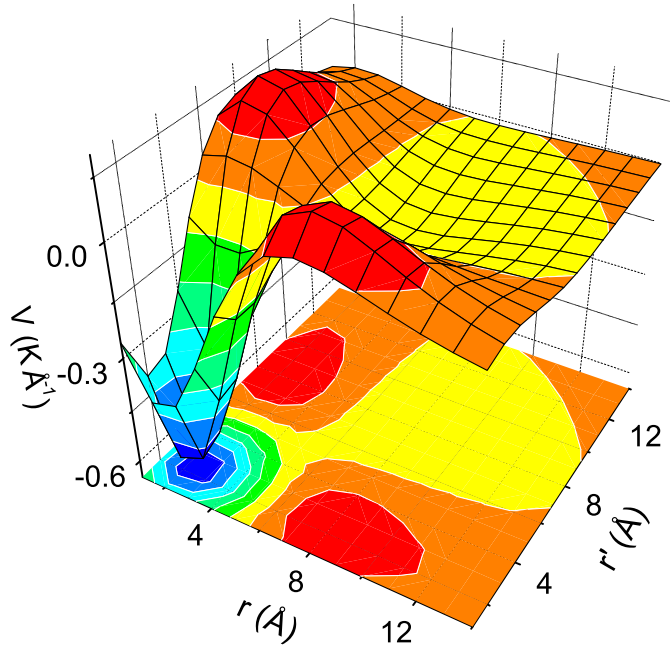}}
\end{center}
\caption{\label{fig:pots}
(Color online)
Configuration space $(r,r')$ representation of the nonlocal rank 2 potentials
in the form $V = r\langle r |v|r' \rangle r'$,
obtained with $\Lambda =0.15$ \AA$^{-1}$ (top) and $\Lambda=0.27$ \AA$^{-1}$ (bottom).}
\end{figure}

Explanation of such a nontrivial behavior of the few-boson system is rather simple.
For this aim one should simply take a look
at the $\Lambda$-evolution of the nonlocal interaction, represented in the configuration space, see Fig.~\ref{fig:pots}.
Potentials with small cutoff $\Lambda$ have single wide and rather flat attractive plateau in the
$(r,r')$ plane. For larger  $\Lambda$
this plateau splits up into a rather complicated saddle-like surface with
two asymmetric attractive regions separated by two symmetric repulsive regions.
One narrow attractive region that deepens with increasing $\Lambda$
is at the origin while wider and flatter region is more distant.
Due to large kinetic energies required to squeeze particles into the narrow region at the origin,
it is energetically preferable for the multi-boson system to reside in the distant attractive plateau.
The resulting spectrum obeys the standard universality relations \cite{Platter,stecher}, 
e.g., it has two tetramer states, one
tightly bound and one weakly bound, which is essentially an atom weakly bound to the trimer in its ground state.
However, at some point close to  $a_0\Lambda = 26.0$ the
 attractive region at the origin becomes deep enough to accommodate the four-boson system, and
some kind of a phase transition occurs. First,
 the excited state of the tetramer accommodates important part of its wave function closer to the origin thereby
loosing its atom-trimer structure and experiencing a drastic reduction of the radius  $R_4^{(1)}$
and a growth of the potential, kinetic and binding energies.
Further deepening of the small $(r,r')$ region brings
the tetramer ground state from the distant plateau to the
attractive region close to the origin. The first excited tetramer state is
pushed away to ``recover the former place'' of the
 ground state, thereby decreasing $K_4^{(1)}$ and increasing $R_4^{(1)}$,
and a second excited state appears in the place of the former excited state. As a consequence,
at  $a_0\Lambda = 27.0$ the tetramer ground state is tightly bound in the narrow deep region close
to the origin while excited states  reside in the  wider flatter region,
the first one being tightly bound and the second one having an atom-trimer structure;
furthermore, their relation to the trimer is nearly universal as for the two tetramer states
in the region $a_0\Lambda < 26.0$.  Thus, at  $a_0\Lambda = 27.0$
it is the  ground state of the tetramer with the
 radius $R_4^{(0)} \approx r_0/5$ much smaller than the range of the interaction
that is anomalous/non-universal one.
 Trimer has only three pair interactions
(compared to six for the tetramer), therefore the corresponding phase transition in the three-boson system
takes place at larger  $\Lambda$ values, when the attractive region near the origin is even more deepened.
As the trimer ground state is brought to the narrow deep region as well, the tetramer states havo to
change correspondingly. Since there is already tightly bound tetramer in that region,
the second excited tetramer state disappears and the first one acquires the atom-trimer structure.
As a consequence, the separation energy $B_4^{(1)} - B_{3}^{(0)}$ decreases while individual binding energies
continue to increase and radii decrease, as can be seen from
the last part of the $R_4^{(1)}$ curve in Fig.~\ref{fig:R}.

Furthermore, we would like to mention that Delfino et al.~\cite{Delfino:92} have found a strong collapse of the
trimer for a rank 1 separable potential with  two-term form factors.
Effect seen in that work may be well related with the one we discuss, since the potential of Ref.~\cite{Delfino:92}
may be rewritten as a rank 2 potential with only two independent $\lambda_{ij}$ parameters and
therefore has the structure comparable to ours with several attractive and repulsive regions. 
However, in  Ref.~\cite{Delfino:92} single two-body parameter, namely $B_2$, was fixed 
and thus collapse could be interpreted as a standard Thomas collapse, being accompanied by the rapid
increase in the average kinetic energy of the dimer (or reduction of the effective range).
On the other hand, our potentials are phase equivalent in a broad energy window, up to $50 \, B_2$, 
thereby constraining in addition two-body scattering length, effective range, and even higher terms 
of the effective-range expansion. Nevertheless, presence of finite effective range
expansion terms ($r_{0}>0$  etc.) does not 
prevent collapse in the multi-boson system. 
Indeed, our additional calculations indicate that varying upper limit of the energy
included in the fit changes only the critical $\Lambda$ values but not the qualitative picture
of the collapse. 

One should probably question the existence of such ``exotic'' nonlocal potentials with two attractive regions.
At present, we cannot give an example of a real physical system where a nontrivial behavior like described
 above is observed. However, we admit that some realistic nucleon-nucleon potentials
\cite{entem:03a,doleschall:04a} have even more
complex structure than that, containing several attractive regions.

\begin{figure}[tbp]
\epsffile{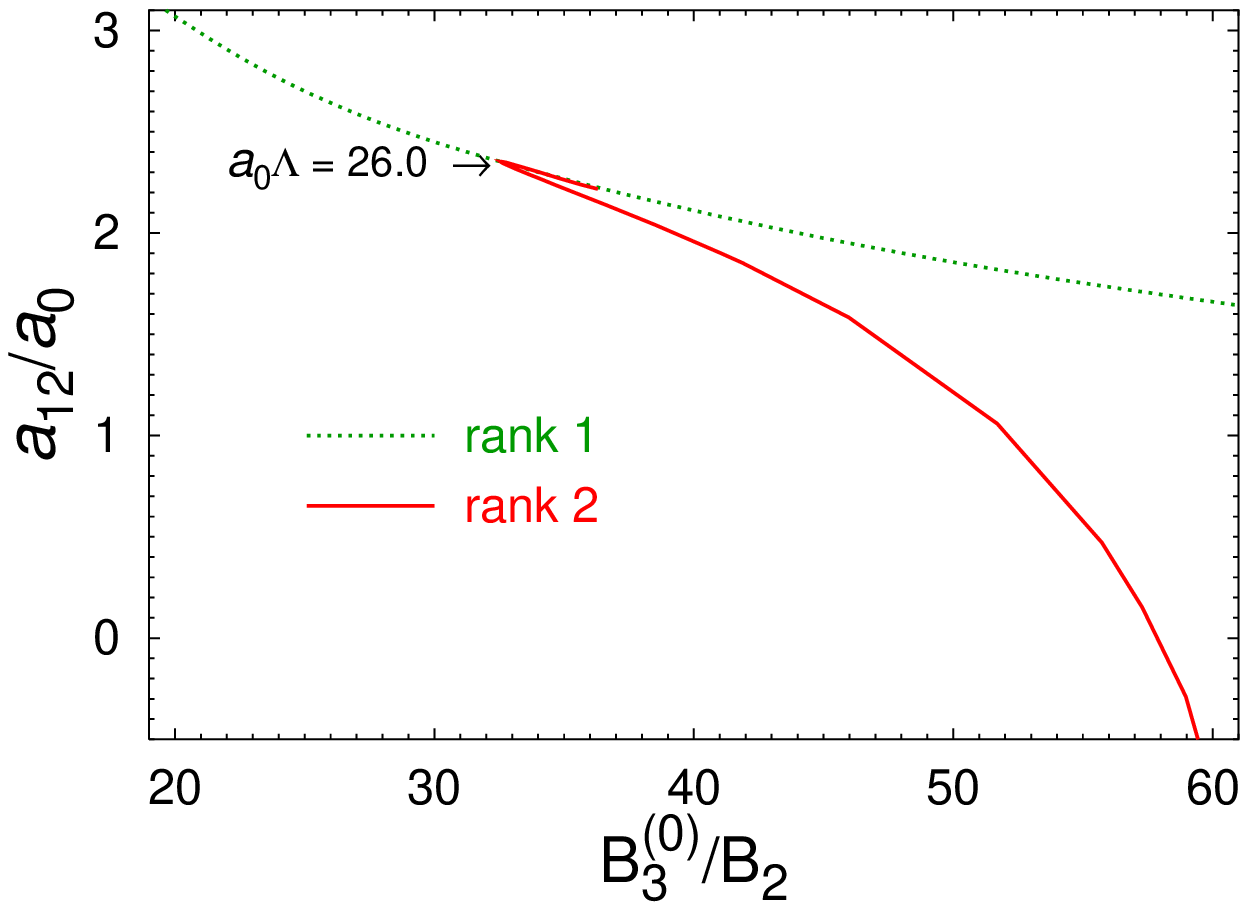}
\caption{\label{fig:ba3}
(Color online)
Correlation between the atom-dimer scattering length and
the trimer binding energy.}
\end{figure}

\begin{figure}[tbp]
\epsffile{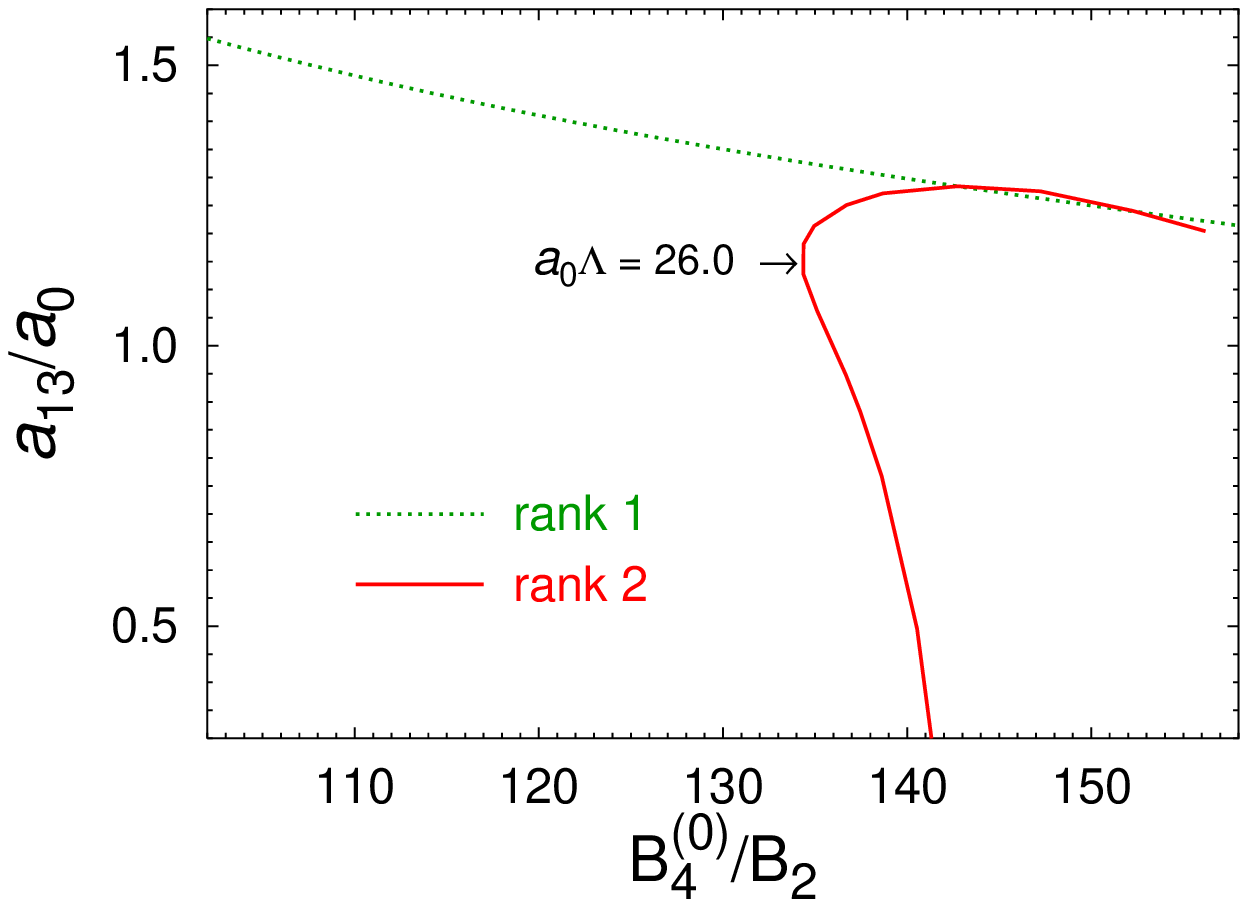}
\caption{\label{fig:ba4}
(Color online)
Correlation between the atom-trimer scattering length and
the tetramer binding energy.}
\end{figure}

Finally, in Figs.~\ref{fig:ba3} and \ref{fig:ba4} we show correlations between the atom-dimer (atom-trimer)
scattering length and the trimer (tetramer) ground state binding energy.
It is nearly linear for rank 1 results, reproducing the well-known Phillips line whereas
rank 2 results agree with that line only up to $a_0\Lambda = 26.0$ or even less in the case of $a_{13}$.
Such a behavior is consistent with previous findings of the present work.

\section{Conclusions \label{sec:con}}

We calculated properties of three- and four-boson systems employing
series of short-range two-boson potentials reproducing the same two-body
binding energy and low-energy scattering observables.
In a large part of the considered parameter space we obtain results that are consistent
with the commonly established EFT results. However, a smooth variation of the potential parameters
leads to the region with drastic deviations form the EFT predictions, e.g.,
rapid variations of tetramer binding energies while trimer binding energies remain almost unchanged
and the appearance of the second excited tetramer state, the reason for these phenomena being
complicated structure of the potential with several attractive regions.

Lately models based on contact interaction EFT have found a huge success in explaining correlations
between the few-particle binding energies in nuclear and cold-atom systems as well as establishing
existence of universal relations between them. The results obtained by us
demonstrate existence of strong deviations from the aforementioned universal behavior.
Indeed, universal behavior is assured only if Efimov condition $r_{0}\ll R\ll | a_{0}| $ is satisfied,
which is fulfilled if one employs contact-interactions (as is used to establish universal relations).
Nevertheless, this condition is not guaranteed a-priori for multi-particle system. Moreover,
the realistic $N>2$ particle systems, even those which are believed to be ideally suited
to apply EFT like cold atomic gases or nuclei, do not always satisfy the first  Efimov condition $r_{0}\ll R$.
For example, $r_{0} \approx R$ for $^{4}$He trimer and tetramer ground states
calculated with ``realistic'' potentials between $^{4}$He atoms \cite{FY_Gre}
or with the present toy model.
The universal behavior still holds in case when the interaction is rather simple, having single attractive
region  and/or $N>2$ and $N=2$ systems are bound to the same attractive region,
however, it is not guaranteed a-priori. A general  $N+1$ particle system may have very different states
from those predicted by the universal EFT, even though all the spectra of $N$ or less particles are well described.

Presence of finite  range ($r_{0}>0$) does not guarantee that Thomas collapse would not occur in multi-boson system.
Interaction can have short-range or off-shell structure that may not be seen with the low-energy probe. Such structure
can have no or little effect for low-energy observables of the systems up to $N$ particles, however may be exploited in the
$N+1$ boson system that,
having more interacting pairs, can compensate larger kinetic energies and thus regroup itself into the shorter range domain.
As discussed in Ref.~\cite{Platter} one can choose three-body scale for repulsive three-body interaction to regularize
 three-body system
and prevent $N$-particles from collapsing. Nevertheless, in special cases
three-body low-energy observables may carry insufficient
amount of information to set the physics beyond $N=3$ system as we demonstrated in the present work and
 a four-body parameter may be needed \cite{Frede_4b}.
 However, this may take place only when the underlying potential has a nontrivial
structure and the size of the few-body system $R < r_0$ becomes smaller than the range of the interaction.
Thus, our results are complementary to the findings  of  Refs.~\cite{Platter,stecher} 
that four-body parameter is not needed under Efimov condition $R >> r_0$ assuring the universality.

\begin{acknowledgments}
The authors thank  A.~C.~Fonseca for the comments on the manuscript.
This work was granted access to the HPC resources of IDRIS under the allocation 2009-i2009056006
made by GENCI (Grand Equipement National de Calcul Intensif). We thank the staff members of the IDRIS for their constant help.
\end{acknowledgments}



\end{document}